\documentclass[twocolumn]{aastex61}
\usepackage{soul}
\usepackage{natbib}
\usepackage{amsmath}
\usepackage{amssymb}
\usepackage{hyperref}
\DeclareMathSymbol{\circledstar}   {\mathbin}{AMSa}{"46}

\bibpunct{(}{)}{;}{a}{}{,}
\usepackage{txfonts}
\def\tens#1{\ensuremath{\mathsf{#1}}}

\newcommand{\A}           {{\tens{A}}}
\newcommand{\M}           {{\tens{M}}}
\newcommand{\F}           {{\mathcal{F}}}

\newcommand{\E}           {{\tens{E}}}

\newcommand{\AWP}         {{\it A-Projection}}
\newcommand{\WBAWP}       {{\it WB A-Projection}}

\newcommand{\ostar}       {{\circledast}}

\begin{document}

\shorttitle{Low Order A-Term Solver}
\shortauthors{Jagannathan~et~al.}

\title{Direction Dependent Corrections in Polarimetric Radio Imaging II: A-Solver Methodology\\
 A low-order solver for the A-Term of the A-Projection algorithm} 

\correspondingauthor{Preshanth Jagannathan}
\email{pjaganna@nrao.edu}

\author{P.~Jagannathan}
\affiliation{National Radio Astronomy Observatory, Socorro, U.S.A}
\affiliation{Inter-University Institute for Data Intensive Astronomy, and \\ Department of Astronomy, University of Cape Town, South Africa.}

\author{S.~Bhatnagar}
\affiliation{National Radio Astronomy Observatory, Socorro, U.S.A}

\author{W.~Brisken}
\affiliation{Long Baseline Observatory, Socorro, U.S.A}

\author{A.~R.~Taylor}
\affiliation{Inter-University Institute for Data Intensive Astronomy, and \\ Department of Astronomy, University of Cape Town, South Africa.}
\affiliation{Department of Physics and Astronomy, University of the Western Cape, Belville, South Africa}


\date{Received: xx-01-2017;  Accepted: xx-01-2017}

\begin{abstract}
  The effects of the antenna far-field power pattern limits the
  imaging performance of modern wide-bandwidth, high-sensitivity
  interferometric radio telescopes. Given a model for the aperture
  illumination pattern (AIP) of the antenna, referred to as the
  A-term, the wide-band (WB) A-Projection algorithm corrects for the effects of
  its time, frequency, and polarization structure. The
  level to which this correction is possible depends how accurately the A-term, represents the true AIP. 
  In this paper, we describe the A-Solver methodology that combines physical
  modeling with optimization to holographic measurements to build an
  accurate model for the AIP. Using a parametrized ray-tracing code as the
  predictor, we solve for the frequency dependence of the antenna
  optics and show that the resulting low-order model for the Karl G. Jansky Very Large Array (VLA)
  antenna captures the dominant frequency-dependent terms.  The
  A-Solver methodology described here is generic and can be adapted for
  other types of antennas as well.  The parameterization is based on
  the physical characteristics of the antenna structure and optics and
  is therefore arguably a compact representation (minimized
  degrees of freedom) of the frequency-dependent structure of the antenna A-term. 
  In this paper, we also show that the parameters derived from A-Solver methodology are expected to improve sensitivity and imaging performance out to the first side-lobe of the antenna.
  \end{abstract}  


%

\section{Introduction}

The aperture illumination pattern (AIP) of an antenna determines the
directional gain and sensitivity of the antenna to the sky brightness
distribution. For an interferometric baseline consisting of a pair of
antennas, the outer convolution of the two AIPs determines the Mueller
matrix.  The Mueller matrix encodes the mixing of the input
polarization signals, including the effects of the off-axis leakage of one
polarization product into another. It also largely determines the
imaging performance.  
Accurate knowledge of the antenna AIP is essential for
high-fidelity imaging performance of a radio interferometric array. 

The current and next generation of interferometric arrays are
outfitted with dual-polarization, wide bandwidth receivers having high
fractional bandwidths (Total bandwidth/center frequency), e.g. in the case of the VLA, by as much as
66--75\% (L, S, and C bands).  The directional properties of the AIP
in each polarization will change significantly across the band.
In addition to smoothly varying geometric frequency scaling of the AIP,
effects can arise due to, imperfect optical alignments or
standing waves between optical elements of the antenna
(e.g. \cite{2008A&A...479..903P}).
Standard calibration and imaging algorithms that do not account for
the directional and frequency dependence of the antenna AIP lead to
errors whose magnitude increases with distance from the antenna
pointing direction and are particularly significant for polarization
imaging \citep{2017paper1}.
With a known AIP, the direction-dependent errors can be corrected over
the field-of-view using the \AWP~algorithm \citep{2008A&A...487..419B,
  2013ApJ...770...91B}. 

The AIP can be modeled to the first order by a simple ray-tracing
geometric model. \textit{Geometric} models of aperture illumination
deliver sufficient accuracy in the regime where the incident
wavelength of electro-magnetic waves is much smaller than the blocking
antenna structures in the optical path. At its highest operation
frequencies of 10s of GHz, the VLA falls in this
\textit{geometric} regime.  However at GHz frequencies and lower
purely geometric approaches are insufficient, higher order effects
from diffraction and scattering significantly affect and alter the
AIP.  textbf{These effects introduce higher order frequency-dependent terms.
Full Electromagnetic (EM) simulations of the antenna \citep{2013ITAP...61.2466Y} in principle
allows for the accurate modeling of the AIP including higher order
effects. However such approaches are
computationally expensive (even more for high spectral resolution
simulation across wide bandwidths) and are limited by the accuracy of
antenna models and illumination patterns which are given as an initial
input. Results from such EM simulations often do not accurately
reflect the real AIPs and are difficult (and expensive) to perturb to
fit the measured AIPs.}

In the forthcoming sections of this paper, we describe a new hybrid method,
called the A-solver, that uses holographic measurements in combination
with low-order parametric modeling of the antennas to efficiently create 
a high spectral resolution model of the full-polarization AIP over the very 
wide-bandwidth of the VLA. We utilize the parameterized beam and using simulations
of point sources across the field quantify the effect of the parameterized AIP on imaging.
The detailed working of the full Mueller A-Projection algorithm and the use of frequency-dependent parameters on imaging of real VLA data will be the focus of a forthcoming third paper.

\subsection{Primary Beam Correction and Imaging}
Corrections for the AIP can be carried out during imaging in the aperture plane(\AWP~algorithm)
 or post deconvolution in the image plane using the Fourier transform of the AIP, the antenna primary beam(PB). 
The PB of radio antennas varies with direction and frequency.  
For altitude-azimuth mounted antennas the sky
brightness distribution rotates with respect to the antenna primary
beam as a function of the antenna parallactic angle. Consequently,
for long integration observations during which the parallactic angle
changes, the response of the array to a radio source includes an
instrumental component that varies with time, frequency and
polarization. In paper-I \citep{2017paper1}, we showed the errors introduced in polarimetric imaging when the time dependence of the antenna PB is unaccounted, in particular, the case of altitude-azimuth (Alt-Az) mounted telescope arrays. Observations with an 
  equatorial mounted antennas or Alt-Az antennas with a third axis of motion to maintain
  a fixed parallactic angle \citep{2016PASA...33...42M} allow for 
  a simple correction in the form of a direction-dependent flux
  subtraction post imaging. This technique was used to good effect for the Canadian Galactic Plane Survey \citep{2003AJ....125.3145T} using the equatorial mount antennas of the Dominion Radio Astronomy Observatory synthesis radio telescope. Alternatively for "snapshot" observations across narrow-bandwidths image plane PB corrections for all polarizations are highly effectively as demonstrated by the NVSS \citep{1998AJ....115.1693C}.
  
At low frequencies where the ionosphere plays a limiting role in full PB direction dependent imaging, peeling based methods (\cite{2008PASP..120..439C} and \cite{2009A&A...501.1185I}) are effective in producing high quality PB corrected images in Stokes I for narrowband surveys like TGSS \citep{2017A&A...598A..78I}. For wide-field dipole arrays such as the MWA, implementations such as the Real-Time System (RTS, \cite{2008ISTSP...2..707M} and \cite{2010PASP..122.1353O}), and WSCLEAN \citep{2014MNRAS.444..606O} allow for image plane PB corrections. Measurements of the polarized MWA PB \citep{2015RaSc...50...52S} provide image plane PB models which are modeled in terms of spherical harmonic functions on the sky \citep{2016icea.confE...4W}. However, small discrepancies in the PB model across wide frequency bands of modern telescopes manifests as a scaling error as a function of declination and frequency, also as noted in the GLEAM survey (\cite{2014PASA...31...45H} and \cite{2017MNRAS.464.1146H}). Polarization observations at low frequencies with the MWA utilize the induced rotation measure by the ionosphere over multiple epochs as a tool in identifying the shifted rotation measure away from $RM=0$ \citep{2016ApJ...830...38L} and \citep{2017arXiv170805799L}. All these approaches require imaging and deconvolution using fractions of data partitioned along all or some of the  axis (time, frequency, polarization, baseline).
 
Aperture plane corrections using wide-band A-Projection algorithm(\cite{2008A&A...487..419B} and \cite{2013ApJ...770...91B}) works on un-partitioned data, thus benefiting from the full sensitivity of modern wide-band telescopes during the non-linear operation of image modeling (a.k.a the "deconvolution" step). \citet{2016AJ....152..124R} demonstrate that multi-scale multi-frequency synthesis (minor cycle) in conjunction with AW-Projection (major cycle) performs significantly better than image plane corrections in joint deconvolution of multi-pointing radio deep fields. In this approach, the modeling can be shown to take advantage of the full available continuum sensitivity of the instrument. With algorithms that require partitioning the data along time or frequency or both, the available SNR for modeling processes is significantly lower. This work therefore, focuses primarily on the full-pol. modeling of the PB for projection algorithms to enable wide-band full-pol imaging including joint-mosaic imaging of complex fields involving a large number of overlapping pointings.

\section{Theory}
\label{sec:SNRtheory}
The measurement equation (ME) for a single interferometer baseline, calibrated
for direction-independent (DI) terms\footnote{All terms that can
 be/are assumed to be constant across the field of view.}, is given by:
\begin{equation}
\vec{V}^{Obs}_{ij}(\nu,t)=W_{ij}(\nu,t)\int
\M_{ij}(\vec{s},\nu,t)\vec{I}(\vec{s},\nu)e^{\iota \vec{b}_{ij}.\vec{s}}d\vec{s}
\label{eq:ME}
\end{equation}
where $\vec{V}^{Obs}_{ij}$ is the visibility measured by a pair of
antennas $i$ and $j$, with a projected separation of $\vec{b}_{ij}$.
$W_{ij}$ are the effective weights, and $\vec{I}(\vec{s},\nu)$ is the
full-polarization vector of the sky brightness distribution as a
function of direction, $\vec{s}$, and $\nu$ is the observing frequency. $M_{ij}$ is the
Mueller matrix which encodes the effects of antenna directional gain
and polarization leakage on the measured visibilities.  $\M_{ij}$ can
be written in terms of the antenna Voltage Pattern (VP), $\E_i$ (following, \citet{1996A&AS..117..137H}), as
\begin{equation}
M_{ij}(\vec{s},\nu,t)=\E_i(\vec{s},\nu,t)\otimes \E^*_j(\vec{s},\nu,t)
\end{equation}

$\M_{ij}$ appears in Eq.~\ref{eq:ME} inside the integral.  Its effects
therefore cannot be calibrated independently of the imaging process to
reconstruct the sky brightness distribution ($\vec{I}$).  They need to
be corrected-for as part of the imaging process using projection
algorithms like \AWP. 
Projection algorithms are a class of radio
interferometric imaging algorithms that correct for the terms inside the
integral of in Eq.~\ref{eq:ME} by applying the inverse of the terms
during convolutional gridding as part of the imaging process
(transforming visibility data to the image domain).  In an
  iterative $\chi^2$-minimization scheme (e.g.,\cite{1995Cornwell} and \cite{2011A&A...532A..71R}),
   the update direction is computed after projecting-out the direction dependen (DD)
  effects, at full accuracy in the
  prediction stage. These algorithms
however, require a model for $\M_{ij}$ as an input, including all the
dominant effects that need to be calibrated.
Equation~\ref{eq:ME} can be recast, in terms of the AIP, which is a
Fourier transform of $\E_i$ as
\begin{eqnarray}
\label{eq:ME2}
\vec{V}^{Obs}_{ij}(\nu,t)&=&W_{ij}(\nu,t)
\mathbf{\mathcal{F}}\left[\left(\E_i(\vec{s},\nu,t)\otimes \E^*_j(\vec{s},\nu,t)\right)\cdot \vec{I}(\vec{s},\nu)\right]\\
&=&W_{ij}(\nu,t) \left[ \A_{ij}\star \vec{V}_{ij}\right]
\end{eqnarray}
where $\mathcal{F}$ is the Fourier transform operator,
$\vec{V_{ij}} = \mathbf{\mathcal{F}}I$ is the true visibility
 full-polarization vector of the sky brightness distribution. $\A_{ij}$ is the Fourier
transform of $\M_{ij}$ and can be decomposed into antenna-based
  quantities as
\begin{equation}
\A_{ij}=\A_i\ostar{\A_j^*}
\label{equation:AIP}
\end{equation}
Here $\A_i$ and $\A_j$ are the AIPs for the two antennas.
Given a model for the AIPs, $\A_{j}^M$, the \AWP\ algorithm
 computes the image as
$\F\left[\A^{M^\dag}_{ij}\star \vec{V}^{Obs}_{ij}\right]$ 
 and the resulting images are normalized by an appropriate function of
$\F[W_{ij}\left(\A_{ij}^{M^\dag}\star\A_{ij}\right)]$ (see
\cite{2013ApJ...770...91B} for details).
$\A^M_{ij}$ is constructed from the models for the AIP of the individual
antennas, $\A^M_i$ and $\A^M_j$ according to Eq.~\ref{equation:AIP}.  
The ability to compute these models accurately and efficiently is 
  therefore crucial for correcting the effects of $\M_{ij}$. 

\subsection{Aperture Illumination Pattern from Holography}
Holography directly measures the antenna VP, $E_i$, 
using the signals from strong unpolarized, compact
calibrator radio source at a grid of positions ($l$, $m$) over the AIP. 
This replaces $I(\vec{s},\nu)$ in Eq.~\ref{eq:ME} with an approximation
of a Kronecker delta function in $\vec{s}$ at each ($l$, $m$).
Typically a subset of the array antennas, whose VP is measured, scan the source, 
while the rest of the antennas are used as reference antennas and are pointed towards the source (i.e. the sources is placed at $l$= $m$ = 0).  
The reference antennas provide the reference signal, with respect to which the 
signal from the scanning antennas is measured, and when projected in the antenna Az-El plane, 
gives a sampled map of the complex VP, $E_i(l,m)$, for all of the scanning antennas.

An important limitation of this method is the low signal in areas of the AIP with 
low directional gain. An accurate aperture model would require the holography measurement to sample 
beyond the first side-lobes in a dense grid with high signal-to-noise ratio. 
Since we are interested in the Fourier transform of the antenna, 
truncation of the measured VP after the first side lobe gives rise to errors 
(due to aliasing) in $\A_i=\mathcal{F}[E_i]$.

$\A_i^{M^\dagger}$ is applied as a convolutional correction while
gridding the observed visibility data onto a regular grid as described
in Eq.~\ref{eq:ME2}.  There are two kinds of oversampling
  required to represent the convolution function (CF) in its appropriate 
  digital form for gridding. For computational efficiency reasons,
  the CF used for gridding in general (not just for
  projection algorithms) is a look-up table. To minimize quantization errors, the CF look-up
  table needs to be oversampled by a factor represented by the
  symbol $O_{ap}$ (typically $O_{ap}\geq20$). Holographic measurements are the measurements of the antenna VP ($E_i$) itself. To minimize aliasing as well as to measure the various
  features of the antenna VP accurately, the holographic measurements
  are also oversampled. However, due to practical limitations, a much smaller oversampling factor ($1.5\times$) was used and found to be sufficient (see Sec. 3.2) for these antenna VP measurements. 
  Since the holographic oversampling factor in the antenna VP measurement is much less
  than the oversampling factor needed for gridding ($O_{ap}$),
  a parameterized model of the antenna AIP is required.

\section{A-Solver: Ray-tracing as a parameterized predictor of the antenna AIP}

\subsection{Physical Modeling of the AIP}
\label{Sec:MODELING}
Approaches to computer models of the AIP or VP can be broadly
classified as Physical Modeling or Phenomenological Modeling.  While a
detailed discussion of these styles of modeling is
 beyond the scope of this paper, we mention here that Physical
Modeling (e.g. simulators using Physical Optics (PO) or full-EM
simulators) minimizes the required degrees-of-freedom in the model and
follows the physics of the problem, both of which have significant
numerical and computational advantages.  Physical Modeling also leads
to a fundamental understanding of the instrument.  Phenomenological
modeling on the other hand\footnote{An extreme example of this is in
  radio interferometric imaging is the Peeling approach where each
  component of the sky brightness distribution is modeled separately,
  parameterized to separately account for each effect, like
  polarization squint, antenna pointing offset, the dependence of beam
  shapes with time, frequency and polarization, amongst others.}  
ignores physics and models individual effects as free parameters.

A simple model of the far-field radiation pattern can be
computed using  geometric optics (GO).  This works well
for smooth surfaces away from edges and structures that are much
smaller than the wavelength of radiation, where diffractive effects
become important.  For observations above several GHz with the VLA,
diffractive effects are expected to be small (e.g. see
\cite{2008A&A...487..419B} where instrumental Stokes-V is modeled
using  a GO simulator for the VP).  
An adaptation of a  GO simulator
\citep{2003VLAMEMO2003} exists in CASA, and referred to as the
``CASSBEAM'' simulator.  The code though VLA centric is general and
can be used to model Cassegrain antennas in general.  The simulator
takes as input a parametric description of the structure of the antenna.  
The VLA antennas are shaped Cassegrain system, with a nearly parabolic primary
and a hyperbolic secondary designed to attain a more uniform
illumination of the antenna aperture. The shaped aperture alters the
side-lobe levels of the antenna far-field, and the side-lobe azimuthal
symmetry is altered by the presence of the quadrupod legs holding up
the secondary. The general shape of the main lobe and the side lobes
is also altered by the central blockage due to the sub-reflector.

\begin{figure}[ht!]
\includegraphics[width=0.5\textwidth]{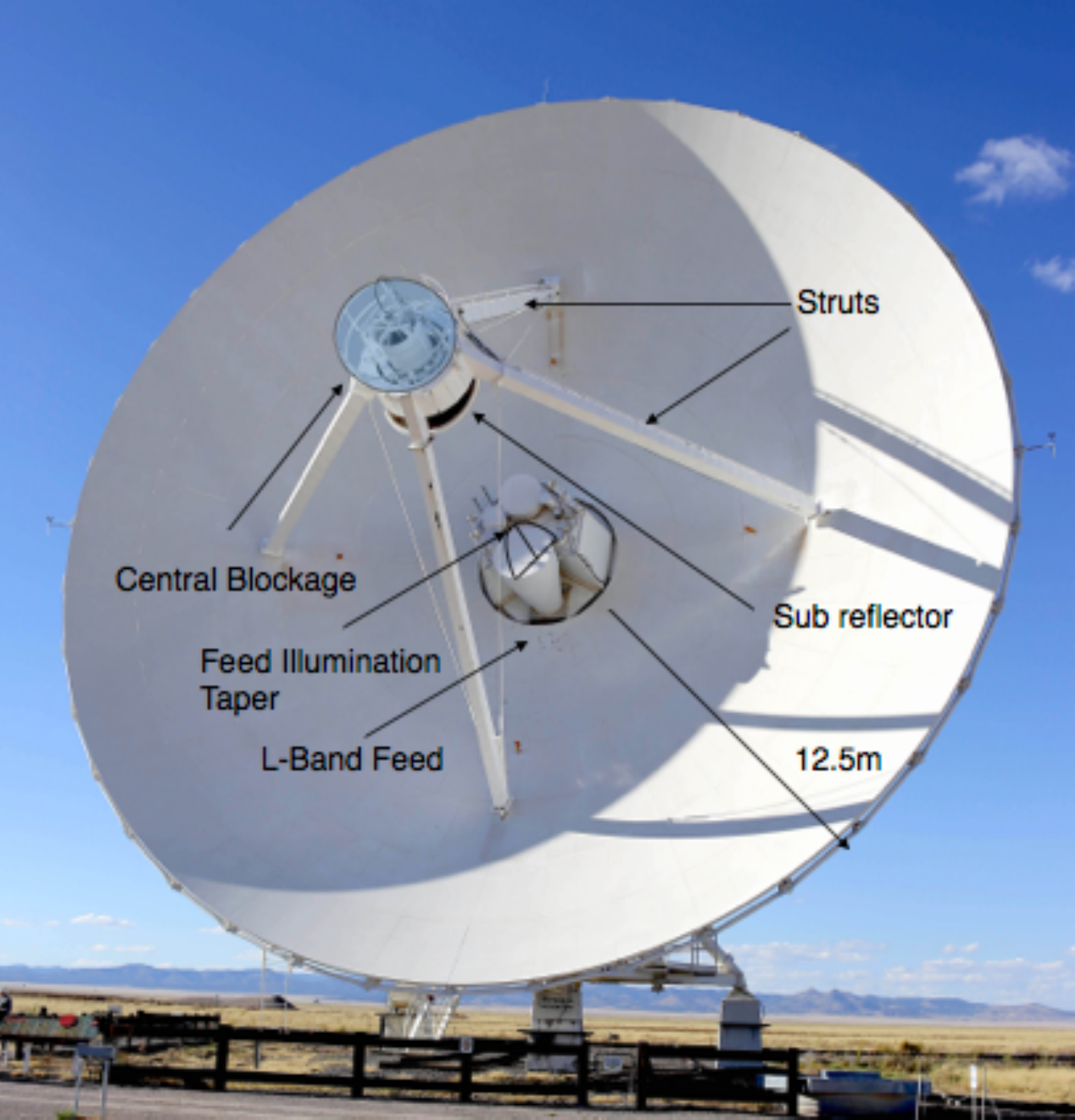}
\caption{ A VLA Antenna is shown with the feeds and the obstructing structs,
along with the two parameters used to modeling the AIP accurately a) the apparent
central blockage (Rhole), b) the feed illumination taper.
The exact physical model parameters used to derive the dish geometry
are described in Table.~\ref{tab:params}}
\label{fig:antenna}
\end{figure}

The set of parameters used in our work to describe the structure and
optics of a VLA antenna are shown in Fig.~\ref{fig:antenna}
are listed in Table~\ref{tab:params}.
The shape of the secondary reflector is not part of the model and is
computed  on-the-fly during ray tracing by
enforcing the optical path length of the rays to be a constant, from
the time of the first incidence. The algorithm computes the changes in the
electric field for the different reflections, following the rays to
the feed where the electric fields in a natural linear polarization basis are
transformed into circular basis having been multiplied by the feed
illumination function and the feed illumination taper function.
CASSBEAM computes all the elements of the direction-dependent antenna
Jones matrix -- the VP for the two orthogonal polarizations along the
diagonal and the leakage patterns on the anti-diagonal, including the
effects of the off-axis location of the feeds (see Fig.~1 of
\cite{2017paper1} and Eq.~3 of \cite{2008A&A...487..419B}).

\begin{deluxetable}{rl}
\tablecaption{Antenna Parameters in Ray Tracing \label{tab:params}}
\tablecolumns{2}
\tablenum{1}
\tablewidth{0pt}
\tablehead{
\colhead{Description} &
\colhead{L Band Values}  \\
}
\startdata
Antenna Name & VLA\\
Sub-reflector height & 8.47852\\
Position of feed in x & -0.10026\\
Position of feed in y & 0.97019\\
Position of feed in z & 1.67640\\
Sub-reflector Angle& 9.26 \\
Width of strut legs & 0.27\\
Strut legs distance from vertex& 7.55\\
Height of strut legs above vertex& 10.93876\\
Radius of central hole & 2.0\\
Radius of the Antenna & 12.5 \\
Band reference frequency & 1.5 \\
Feed taper polynomial  & 10.0, 2.0\\ 
Order of feed taper polynomial & 2 \\
\enddata
\tablecomments{All measurements of length are in meters. All angle measures have units of degrees. All frequencies are in GHz. Polynomial coefficients are unit less quantities. All dimensions provided here are from \citep{memo5}}
\end{deluxetable}
This simple \textit{geometric} model of the antenna aperture illumination
 is insufficient for \AWP.  At L, S and C Bands (1-2, 2-4, 4-8 GHz) of the VLA, 
secondary reflections, diffraction, and
scattering play a major role while standing-waves in the optics play
a significant role in all the bands. Secondary scatterings involving the
feed, sub-reflector, support beams and struts -- all structures of
the order of the wavelength of incident electromagnetic waves -- alters the PB. 
The effects on the antenna PB are two-fold -- the amount of flux gets redistributed 
 from the main lobe to the side-lobes, and the introduction of higher order
frequency-dependent effects in the antenna PB, that alters
the effective off-axis leakage, and the angle of polarization squint.

To refine the AIP model we developed the A-solver approach where we 
perturbed the model parameters such that the predicted AIP fits the 
Holographic measurement of the AIP.  The latter is a measure 
of the real AIP, which usually is significantly different from idealized
AIP.

\subsection{Holography}
\label{sec:HOLO}
For the holography data used here, an unpolarized
source, 3C147 ($<0.04$\% polarized), was scanned in a $35\times35$ grid
in the antenna reference frame with a step size of $\Delta l, \Delta m = 2.5057\arcmin$,
 out to the second null (at 1GHz). In addition, a polarized calibrator 3C286 was observed to provide
polarization angle calibration. Half the array was utilized as
reference antennas 
while the other target antennas scanned the array. For more details on
the holographic measurement see \cite{2016VLAMEMO}.

The visibility data were imported into AIPS to obtain the antenna grid
coordinates ($l$,$m$) for each holography scan. The uncalibrated data
were exported as UVFITS file and imported into CASA as a Measurement Set. 
The calibrator fluxes were set using \cite{2013ApJS..204...19P}
and \cite{2013ApJS..206...16P} for both 3C147 and 3C286 in all
polarizations across the full bandwidth. On-axis gain, bandpass, frequency-dependent 
polarization leakage and polarization position angle
calibration  were carried out and the calibration solutions were
applied to the data using the {\tt APPLYCAL} task in
CASA. Subsequently, utilizing the CASA toolkit,  data from baselines
between the target antennas and each of the reference antennas were averaged to improve
the signal-to-noise ratio of the measured VPs. Further, the data recorded
on a grid point for 10 seconds was averaged. This gave the final set of
antenna VP data per channel per holography grid.  The VP data were
then interpolated onto a $128\times128$ grid on a per channel basis to
create a 1024-channel image cube for each of the target antennas, in
polarizations {\tt R} and {\tt L} (the diagonal elements of the DD
antenna Jones matrix) and leakage patterns {\tt R$\leftarrow$L}, {\tt
L$\leftarrow$R} (the anti-diagonal elements of DD antenna Jones
matrix).

\subsection{A-solver Optimization procedure}
\label{sec:optimization}
The ray-tracing AIP simulator code within the CASA imaging R\&D code
base was modified to accept input parameters from a python wrapper
code. This was wrapped as a parameterized function in Python and
utilized as the unknown function to be determined by the Nelder-Mead
simplex algorithm \citep{1965nelder}, minimizing for the function
parameters against each of the individual channel images of the target
antenna VP cube produced from the holography data (see
Section~\ref{sec:HOLO}). The optimization parameters chosen were the apparent
blockage (Rhole), the feed illumination taper function (ftaper as a $4^{th}$ order
polynomial), and the antenna pointing offset in  R.A. and Dec (xoffset,
yoffset)\footnote{The antenna pointing offset we solved for is the
  mechanical antenna pointing, different from pointing offsets due to
  polarization squint. The latter is an optical phenomenon and is
  naturally included in the computation of the VP due to the physics
  of off-axis optics and does not need to be solved-for as an
  independent parameter}. The apparent blockage
  parameter along with the feed illumination taper function altered the antenna AIP and
  consequently the antenna PB, without altering the optical path of the incident radiation.
  While these parameters appear independent they are not orthogonal and produce the best antenna AIP for
  a given frequency together. An initial run including only the apparent blockage and feed illumination 
  taper gave higher systematic gradients. Such gradients signify physical antenna pointing errors, which were independently parameterized and included as part of the optimization procedure.
 The simplex algorithm traversed each
parameter space independent of the others varying them until a joint
minima is found. The residuals before and after the joint minimization
are shown in the upper and lower panels of
Fig.~\ref{fig:chan_compare}. The choice of the simplex algorithm over
more computationally optimal algorithms arises from the lack of
\textit{apriori} knowledge of the gradients of the various
minimization parameters, in the seven-dimensional parameter space.

The CASSBEAM code uses OpenMP thread parallelization and was
set to launch four threads per process call. This parallelization allowed for
the fast production of a new beam model for every convergence
iteration.  Despite this parallelization the minimization takes 4 hours
per channel per polarization to converge to a solution. So a serial
minimization would take $2\times 1024\times 4$ hours to derive a
channelized solution for an antenna. Since each channel minimization
based on our parameterization is independent of the next channel a
simple frequency-based parallelization was used to trigger a parallel
minimization run of 1024 channels and two polarizations on the Amazon
Web Services (AWS) compute cluster.  This reduced the compute time down to 6
hours per antenna.  Our results for three antennas are
discussed below.  We should note here that the run-time would be
unreasonably long if a full Physical Optics simulator like GRASP \footnote{\url{http://www.ticra.com/products/software/grasp}} was used instead.

\section{Results}

\begin{figure*}[ht!]
\includegraphics[width=\textwidth]{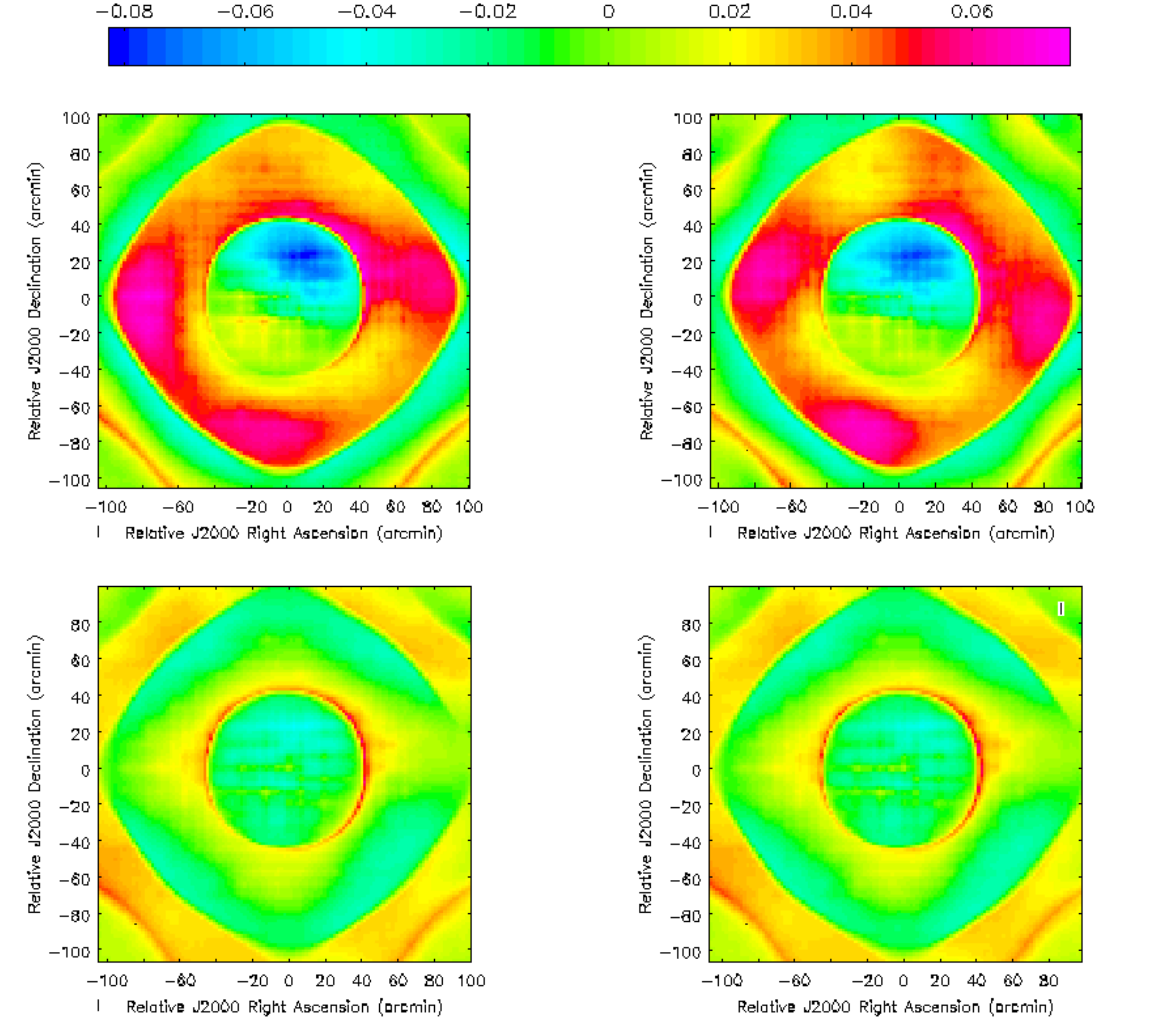}
\caption{The upper panel images shows the residuals (normalized with respect to peak intensity at the beam center) of $|E_{i}^{Holo}
  - \mathcal{F}^{-1}A_{i}^{ideal}|$ measured at $1.353$~GHz. The upper
  left panel show the residual for the left-circular ({\tt L}) polarization
   and the upper right panel for the right-circular ({\tt R}). 
   Similarly, the lower panels show $|E_{i}^{Holo} - \mathcal{F}^{-1}A_{i}^{M}|$. }
\label{fig:chan_compare}
\end{figure*}

In order to highlight the efficacy of the parameterized model of
$A_{i}^{M}$ as against the \textit{ideal} model of the AIP's 
Fig.~\ref{fig:chan_compare} shows a comparison between
$|E_{i}^{Holo} - \mathcal{F}^{-1}A_{i}^{ideal}|$ (top row) and the
$|E_{i}^{Holo} - \mathcal{F}^{-1}A_{i}^{M}|$ (bottom row).
 $A_{i}^{ideal}$ refers to the default aperture illumination
produced by the CASSBEAM code, and $A_{i}^{M}$ is derived from the
parameter values obtained by the optimization procedure discussed in
Sec.~\ref{sec:optimization}. The first side-lobe
 is underestimated in the upper panels of
 Fig.~\ref{fig:chan_compare} by 50\% in both polarizations ({\tt L}
upper left and {\tt R} upper right). Within the main-lobe of the VP,
the residuals in the upper panel show
a systematic offset in power within the main-lobe in both
polarizations. The offset within the main-lobes that affects both
polarizations equally is a sign of mechanical antenna pointing
error.  In contrast, the lower panel images (lower-left for
{\tt L} lower left and lower right for {\tt R} lower right) of the parameterized model residuals shows
no sign of side-lobe power discrepancy or residual pointing error. 
These are residuals for one frequency channel.
The optimized residuals show similar improvement across the entire bandwidth at the VLA L-Band 
for all the optimized antennas. 

The optimization procedure solved for the pointing offset of the
antenna and then fitted the data for \textit{Rhole}
 -- the \textit{apparent} blockage. \cite {2001PASP..113.1247H}
demonstrated that a blocked aperture leads to increased power in the
side-lobes. They also show that the size and extent of the VP
side-lobes can be effectively shaped by tapering the illumination of
the feeds. In line with their finding we were able to effectively
model the first side-lobe power altering the \textit{apparent}
blockage in ray-tracing in conjunction with the \textit{ftaper}, feed
illumination taper polynomial function utilized in our code. We find
that an increased apparent blockage and a sharper tapering function
for feed illumination, determined per channel across the entire band
allows for the capture of all significant changes in the antenna VP
out the first side-lobe. We also note that the
trend captured in the optimized parameters correlates with the
measured wide-band sensitivity of the VLA L-Band \citep{memo165},
which suggests that our optimized models correctly estimates the departures
from idealized antenna.

\subsection{Apparent Central Blockage}
\label{sec:ACB}
\begin{figure}
\includegraphics[width=0.5\textwidth]{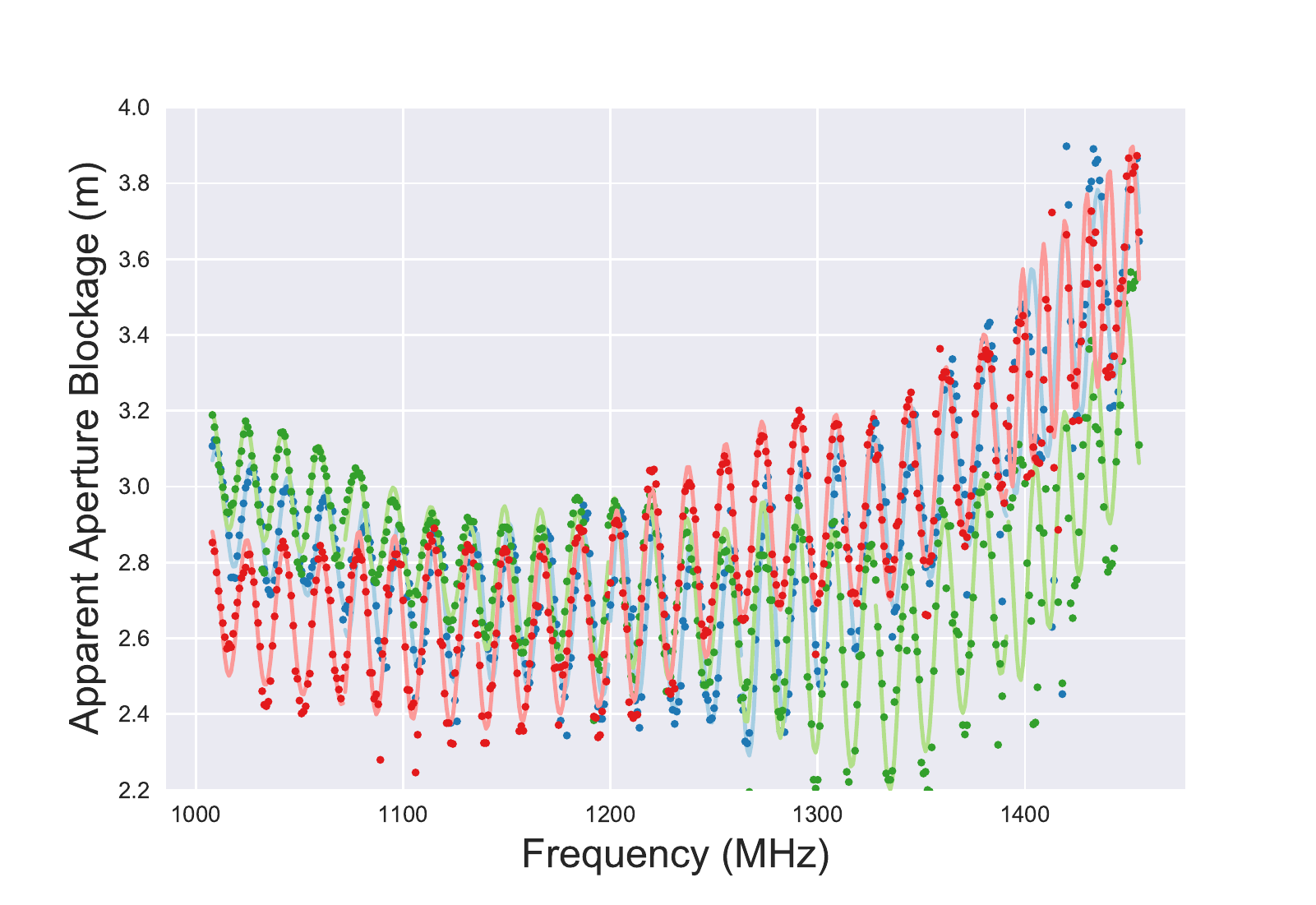}
\caption{Fit to the recovered \textit{apparent} blockage parameter for antennas 6, 10, 12, in red, green and blue respectively, with the lines representing the fit and the points represent the derived apparent blockage data across $448$~MHz, of data.}
\label{fig:Rhole_fit}
\end{figure}

The central blockage in an antenna reduces the aperture efficiency and
increases the side-lobe levels -- an aspect that is
alleviated by shaped surface design and off-axis feed geometry of the
VLA to improve uniform aperture illumination and increased aperture
efficiency (chapter 3, \cite{1999ASPC..180.....T} ). The frequency
dependence of the VP across the bandwidth, in particular, the presence
of a standing wave, altering the first side-lobe and the
shape of the polarization properties of the VP for the 
JVLA antenna across L-Band. Solving for a frequency-dependent
apparent blockage allowed us to capture the frequency-dependent
variation in the per-channel solutions of the \textit{Rhole}
parameter. Plotted in Fig.~\ref{fig:Rhole_fit} in red, green and blue
is the \textit{apparent} blockage parameter for three different
  antennas derived from the optimization spanning seven
spectral windows.  The effect of the
standing wave is captured in the variation of the
\textit{Rhole} parameter with frequency.  This frequency-dependent
variation of the \textit{Rhole} parameter, in turn, can be fit
using a combination of a straight line in frequency per spectral
window, and a sinusoidal function.
The data are fit per spectral window each containing $64$~MHz of data
utilizing the Astropy models package \citep{2013A&A...558A..33A}. The
fit reveals that the oscillations in frequency have a
period of $\sim~17$~MHz. The period of this
oscillation corresponds to twice the light travel time from the feed
to the secondary, consistent with the presence of a standing wave between
the antenna secondary and the feed. With this fit -- of a line and a
sine function -- the number of parameters that
determine the frequency-dependent behavior of the antenna AIP 
 to five numbers per spectral window. (The data and the fits
to the data are available upon request). The standing waves 
in the apparent blockage is a static effect, that arises 
from a second reflection between the feed and the antenna secondary. While
these static effects are common to all the antennas analyzed, there were differences
in the average trend per frequency from one antenna to another. These antenna-to-antenna variations can be naturally accounted for in the general A-Projection framework. 
The variations in the parameter could be from differences in the optics from antenna to
antenna where small difference lead to measurable differences in the antenna PB. 

\subsection{Feed Illumination Taper}
\label{sec:FIT}
\begin{figure}
\includegraphics[width=0.5\textwidth]{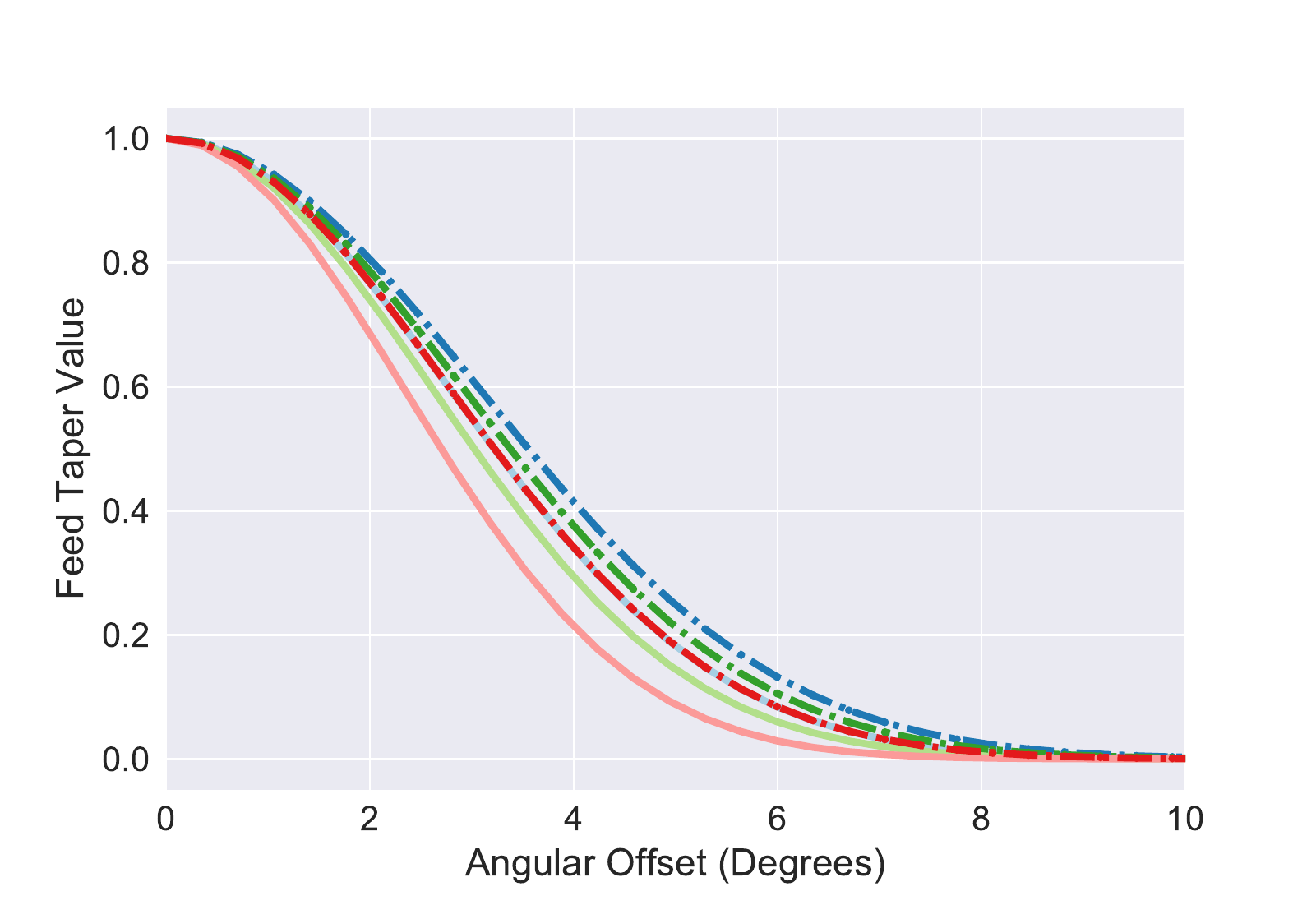}
\caption{The dashed red, green and blue lines show the feed taper function at 1, 1.5 and 2.0~GHz respectively, used to derive $A_{i}^{ideal}$. The solid, red, green and blue (overwritten by the dashed red line) lines show the feed taper function at 1, 1.5 and 2.0~GHz respectively, used to derive $A_{i}^{M}$ for antenna 12.}
\label{fig:feedtaper}
\end{figure}

The VLA receiver feeds lie on a circle around the optical
axis. The feeds are illuminated by the sub-reflector and the angular
span of the illumination can be altered by tapering the feed
illumination pattern. The tapered illumination pattern reduces the
amount of radiation received from the edges of the dish, which while
marginally reduces aperture efficiency, effectively stops the feed
receiving spillover radiation.  In addition to reducing the spillover,
it also alters the shape and the gain of the
PB side-lobe. The parameterized AIP
model allowed for the taper function (a $4^{th}$~order polynomial) to
vary along with the central blockage to optimally match
the shape and structure of the antenna VP out to the first-lobe. In
Fig.~\ref{fig:feedtaper} the normalized amplitude of the feed taper
function is plotted against the angular distance from the feed axis
for antenna 12, at 1.0, 1.5 and 2.0~GHz in red, green and blue dashed
lines respectively. The feed taper function obtained from the
optimization is plotted in light red, green and blue solid lines
respectively. The taper functions determined from parameter
optimization have stronger tapering and a sharper fall off resulting
in lesser feed illumination overall to match the VP's determined
through holography.

\subsection{Pointing Offset}

\begin{figure}
\includegraphics[width=0.5\textwidth]{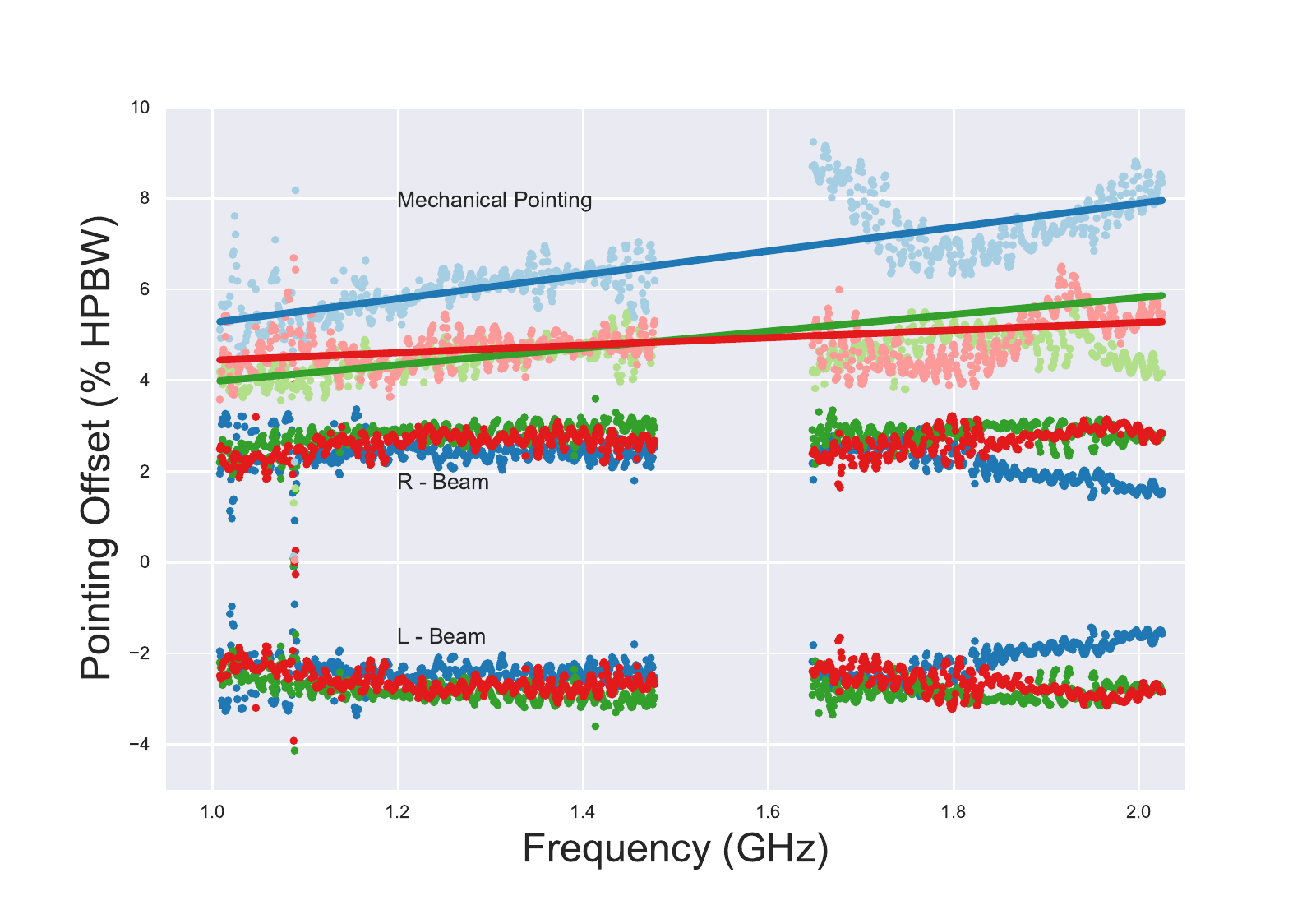}
\caption{Plotted are the pointing offsets for antennas 6, 10, 12, in blue, green and red respectively, for {\tt R}  and {\tt L} polarizations of the antennas around zero. The offsets are in percentage of the HPBW of the antenna. The lines represent the actual pointing vectors of the antennas and is obtained by fitting the per channel pointing solutions with a best fit line. The regions with no solutions between 1.4 and 1.65 GHz corresponds to two spectral windows with noisy solutions due to the presence of strong radio frequency interference. }
\label{fig:pointing_offset}
\end{figure}

Fig.~\ref{fig:pointing_offset} plots the per-channel solutions for the
pointing offsets for antennas 6, 10 and, 12 in
blue, green, and red, respectively in units of the half-power-beam-width
(HPBW).  Any linear scaling with frequency is therefore removed
  and all optical effects that scale linearly with frequency should
  appear as flat curves in this plot.  On the other hand, effects like
  the mechanical pointing offsets, which are not optical effects,
  should appear in this plot with linear slope as a function of
  frequency.  The mean separation between the R- and L-beams is
  $\sim5.7\%$ corresponding to the known polarization squint due to the
  off-axis optics of the JVLA antenna. The solid lines and the
fainter points plotted above the curves showing the R- and L-beam
  offsets are the  mechanical pointing offsets for the three antennas.  Antenna
6 shows the largest pointing offset indicated by the
  the line-fit with a slope of $\sim~2.4\arcmin$. All antennas show mild
  variations in the pointing offsets with frequency. Frequency
dependent pointing error over and above the squint can be caused by an
uncorrected second order term in phase across the antenna. The higher
order phase terms also affect the $ E^{R \leftarrow L}$ and $E^{
  L \leftarrow R}$ adversely, introducing squash and other higher order
distortions.  Modeling these higher order phase errors in the
off-diagonal Jones matrix is covered in Sec.~\ref{sec:squash}.  Once
the pointing offset has been solved-for per channel, solutions for the
\textit{apparent} blockage (Rhole) and the feed illumination taper
polynomial (ftaper) were derived.

\begin{figure*}[ht!]
\includegraphics[width=0.95\textwidth]{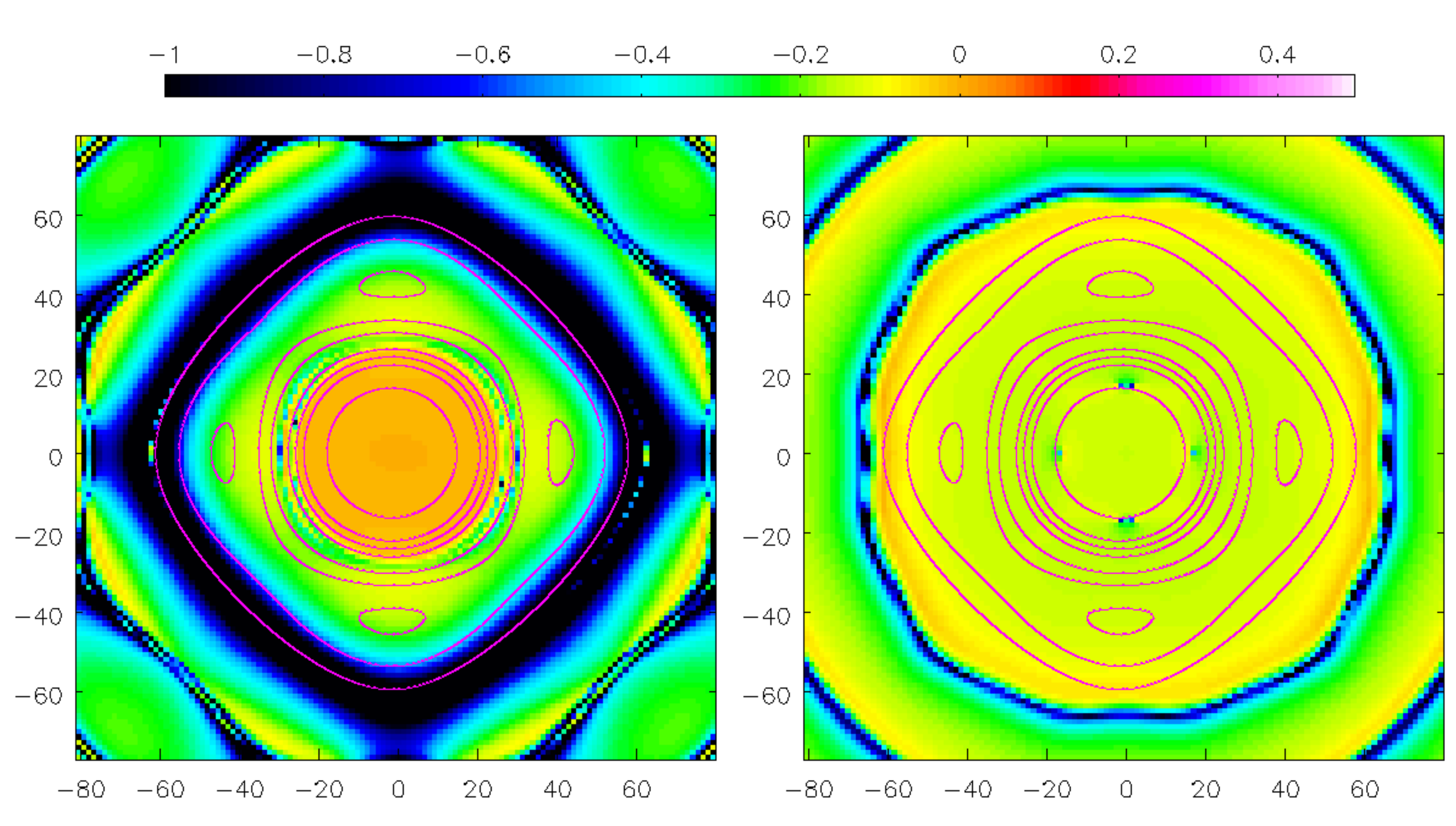}
\caption{ Plotted is the fractional change in the antenna PB, $\left( PB^{def} - PB^{True}\right)/ PB^{True}$, with magenta contours overlaid of $PB^{True}$ at 80, 50, 10, 5 and 1 \% power at 1.448~GHz of antenna 12. The left panel is the fractional change in total intensity, while the panel on the right is the fractional change in linear polarized intensity}
\label{fig:fracpol_pb}
\end{figure*}

\subsection{Antenna AIP and Imaging}
\label{sec:AIP}
A sub-optimal AIP model,
$\A_{ij}^{M}$, will create errors in the image that can be
characterized in the residual image.  The residual
error contribution in a snapshot for a single baseline
$i--j$ can be written as,
\begin{equation}
I^{res} = I^{psf} \star[ \Delta M_{ij} \cdot I^{\circ}]
\label{eq:error}
\end{equation}
where $I^{res}$ is the residual image, $I^{psf}$ is the 
telescope point spread function to be deconvolved, $I^{\circ}$ is the true sky distribution, $\Delta M_{ij} = \mathcal{F}[A_{ij}^{True}] - \mathcal{F}[A_{M}]$ is the difference between the true antenna AIP and the model AIP.

Let us consider $PB^{True}$(or equivalently
$\mathcal{F}[A_{ij}^{True}]$) to denote the PB of the antenna AIP with
optimized \textit{Rhole} and \textit{ftaper} parameters, and
$PB^{def}$ to denote the PB of the antenna AIP with frequency
independent \textit{Rhole} and \textit{ftaper}
parameters. The left panel of Fig~\ref{fig:fracpol_pb} then shows the fractional error
$(PB^{True} - PB^{def})/PB^{True}$ when using the standard sub optimal
AIP, as against the optimized AIP, for stokes I at 1.448~GHz of
antenna 12. The optimized beam is overlaid as contours in pink. The
error within the main lobe of the PB is at the level of several percents, 
a significant change for high fidelity imaging noise limited 
wide-field imaging that typically requiring dynamic ranges in excess of 10,000:1.
The left panel of Fig.~\ref{fig:fracpol_pb} also demonstrates that error in flux
reconstruction $> 5\%$ start beyond the 0.05 gain position of the PB
main lobe and continues to increase to nearly 40 -- 60\% change across
the first side-lobe. 
At right in Fig~\ref{fig:fracpol_pb} is the
fractional error in polarized intensity $(PB^{True} - PB^{def})/PB^{True}$.
The error in the polarized intensity varies between 10 and 20\% across the 
PB out to the first side-lobe.

\begin{figure}[ht!]
\includegraphics[width=0.5\textwidth]{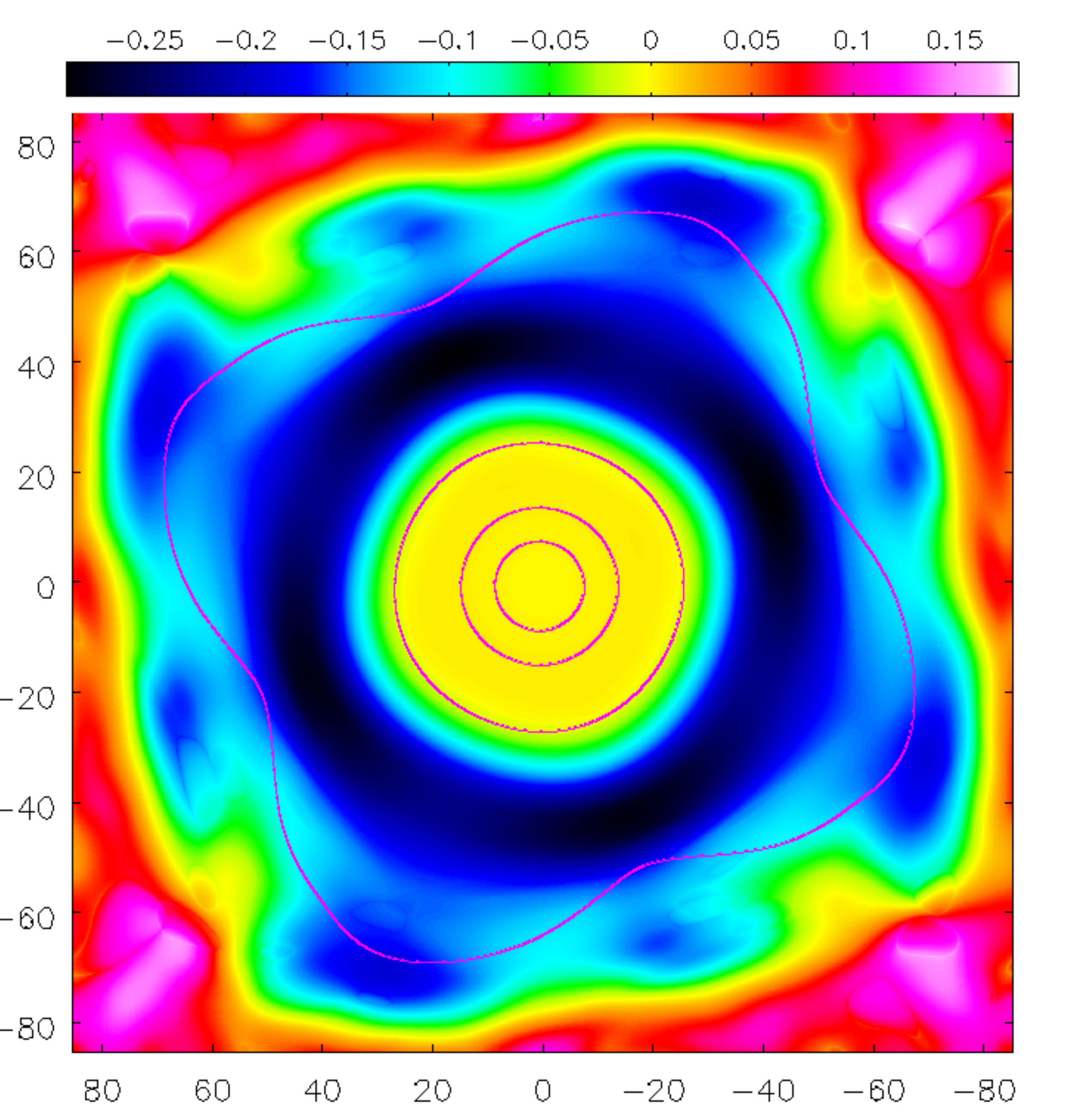}
\caption{ Plotted is the fractional change in the antenna wideband PB, $\left( PB^{True} - PB^{def}\right) / {PB}^{True}$ across 1~GHz of bandwidth, with magenta contours overlaid of $PB^{True}$ at 80, 50, 10, and 1 \% power at the reference frequency,1.5~GHz of antenna 12.}
\label{fig:frac_pb_wb}
\end{figure}

While the fractional error in the PB gives us the
instantaneous error in the residual image for a particular frequency
the effect on the total continuum sensitivity offered by the wide
bandwidths is gotten by examining the fractional error in the
wide-band PB.
The instantaneous wide-band PB is defined as $\sum_{\nu_0}^{\nu1}
PB(\nu)$ spanning the range of frequencies from $\nu_0$ to
$\nu_1$. The wide-band PB represents the effective forward gain 
of broad-band continuum imaging. The effective wide-band sensitivity extends far beyond the null 
of the narrow band PB \citep{2011ApJ...739L..20B}. 
\WBAWP~uses the wide-band PB to normalize the image in the final
imaging step of the \textit{flat-noise} implementation of the
algorithm (see \cite{2013ApJ...770...91B} for more details). Shown in
Fig.~\ref{fig:frac_pb_wb} is the fractional error in the wide-band PB
at the reference frequency of 1.5~GHz. Overlaid in pink are the
contours of instantaneous wide-band $\sum PB^{True}$. The fractional
error in the PB means the error in gain of the
PB-corrected image is $\sim 5\%$ at the 0.1 gain of the PB and
increases to  $\sim 20\%$ at 0.01 PB gain (this includes the
  first side-lobe). Since every pixel in the
  wide-band PB image is the sum of the pixel values at all the
  frequencies, the fractional error beyond 0.1 PB gain is dominated by
  the lower frequencies (larger beam size)  while the error within the 0.1 PB gain being dominated by the higher frequencies. 

\subsection{Imaging Simulations}
\label{sec:simulations}

We used point source simulations to contrast the difference  between parameterized AIP and frequency-independent model for full Mueller imaging.Eight unpolarized point sources (I=1Jy, Q, U, V=0), were placed across the main-lobe and first side-lobe of the antenna PB. The data were simulated for a total integration time of fifteen minutes, with a bandwidth of 64MHz centered at 1.4GHz to produce a full Mueller predicted measurement set (MS)(refer fig 4, \cite{2017paper1} for schematic) for the VLA in C-Configuration. The median value of the apparent central blockage (refer, sec~\ref{sec:ACB}) and feed illumination taper (refer, sec.~\ref{sec:FIT}) of antennas 6,10, and 12 were used as inputs to CASSBEAM. The resulting Jones matrix was used as an in put in our simulations.

\begin{figure}[ht!]
\includegraphics[width=0.5\textwidth]{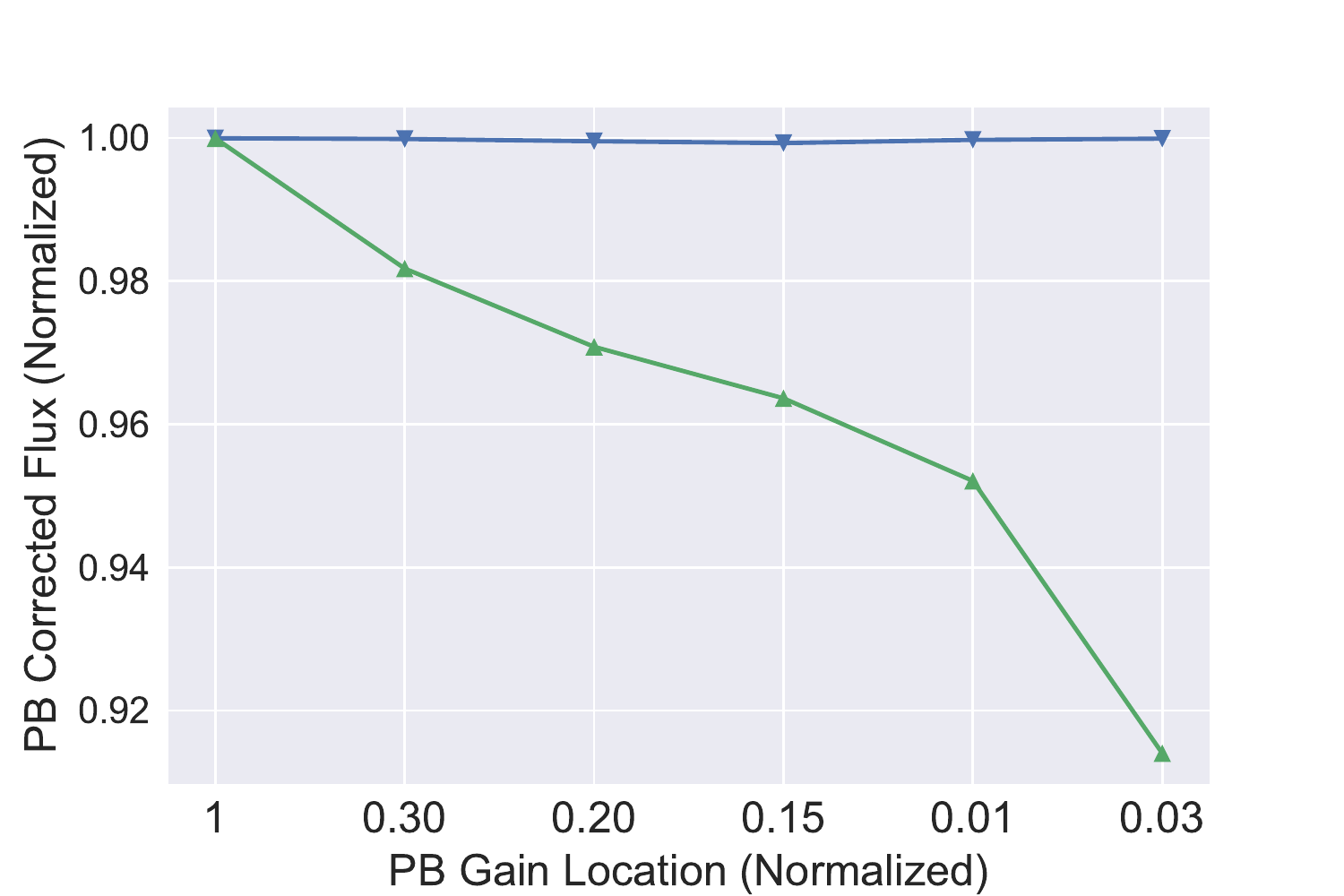}
\caption{Plotted in the figure is the PB corrected point source flux. Plotted in blue are the full mueller imaged and $PB^{True}$ corrected point source fluxes for the parameterized frequency-dependent model. Plotted in green is the full mueller imaged and $PB^{def}$ corrected point source fluxes for the frequency independent model.}
\label{fig:point_source_flux_compare}
\end{figure}

The MS was then imaged with full Mueller A-Projection with the convolution functions produced with a) frequency independent (default) parameters for the feed illumination taper and central blockage, and b)with the updated (frequency-dependent) parameters. We refer to the PB derived from default parameters as as $PB^{True}$. The reconstructed fluxes as a function of the PB gain is shown in Fig.~\ref{fig:point_source_flux_compare}. The blue curve (using the optimized parameters, $PB^{True}$) shows that we are able to reconstruct the flux in total intensity accurately when utilizing an accurate frequency-dependent AIP in the full Mueller A-Projection algorithm. The green curve (standard parameters, $PB^{def}$) shows that when using a frequency independent AIP pattern we begin to incur errors that increase from $\sim 2$\% at the 0.3 gain in the PB to $\sim 9$\% at the 0.01 gain within the main-lobe. In addition to the six sources in the main-lobe two more source were places in the side-lobes, where the standard parameters over-estimates flux by $\sim 25$\%, as we divide by $PB^{True}$, which underestimates the power in the side-lobes.

\begin{figure}[ht!]
\includegraphics[width=0.5\textwidth]{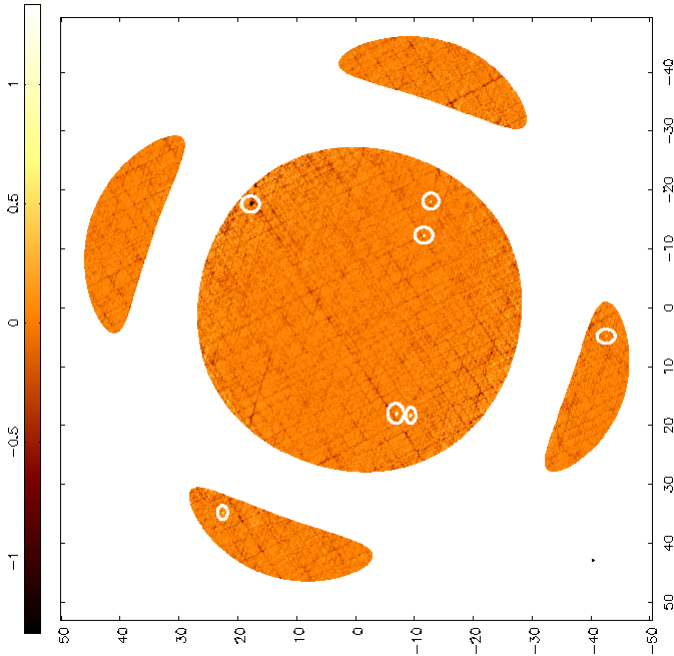}
\caption{The difference image $I^{True} - I^{def}$. The white circles mark the locations of the points sources in the image. The color scales from $-1\times 10^{-4}Jy/beam$ to $1\times 10^{-4} Jy/beam$.}
\label{fig:diff_point_source}
\end{figure}

Fig.~\ref{fig:diff_point_source} shows the difference image, $I^{True} - I^{def}$, with the point source locations indicated with white circles. The color scale is chosen to highlight the deconvolution errors introduced. These errors are more prominent beyond the 5\% PB gain mark within the main-lobe. The deconvolution errors denote the loss in fidelity of imaging and represent degradation in imaging fidelity, even though the effects are markedly visible only when the imaging dynamic range is in excess of 10000:1.

Note that the A-Projection framework used for imaging naturally includes antenna to antenna 
variations, in particular, to account for the AIP of heterogenous arrays such as ALMA. In this paper
we have modeled the dominant static term of the antenna AIP in terms of the feed illumination taper
and the apparent central blockage parameters. While the pointing offset we solve for is used to derive
a better fit to the antenna AIP we are aware that it is a time varying quantity that as a part of the A-Solver approach cannot be described in this paper. Time dependent pointing effects however, can be solved-for by
means of the Pointing Selfcal approach(\cite{2004VLAMEMO}, \cite{2017pointing}). Time dependent shape changes that affect 
the  antenna AIP derived from the A-Solver methodology would only affect the highest dynamic range 
imaging studies ($\geq10^{6}$) for homogenous arrays. In which case a coupled shape and pointing 
self-calibration approach would be required.

A few computational points of note with respect to the full Mueller A-Projection (FM-AW)P framework are worth mentioning at present. A more detailed presentation of the algorithm and its performance on observations will be presented in a forthcoming paper. The CF production in FM-AWP as implemented in CASA, is on a per spectral window basis (typically 16 to 32 spectral windows across a VLA observing band). The CFs are produced once at the start of the imaging and cached. In a typical A-Projection imaging cycle $~80\%$ of the time is spent in the gridding of the data. The convolution function production even is a significantly lesser fraction, typically $~10\%$ of the total imaging time and is a one time cost as the convolution functions are cached. Within the new imager framework convolution function production and gridding are parallelized\footnote{\url{http://www.aoc.nrao.edu/~sbhatnag/misc/Imager_Parallelization.pdf}} by means of MPI. This parallel framework has made the FM-AWP algorithm computationally feasible.

\subsection{Off-Diagonal Antenna Jones}
\label{sec:squash}
In paper I \citep{2017paper1}, we demonstrated the effects of beam \textit{squash} on polarimetric imaging. \textit{Squash} is caused by a second order phase term \citep{2001PASP..113.1247H} and in conjunction with other second order phase terms like defocus and coma, affect polarimetric imaging adversely. Reconstructing the polarized emission from the sky requires the use of all the terms of the antenna Jones matrix. In the 
prior sections, we have dealt with the frequency dependence of the antenna AIP primarily in the context of 
the diagonal elements of the antenna Jones matrix. To model the off-diagonal jones elements requires the  inclusion of higher order distortions which was done by including a general second order polynomial 
in phase in addition to the \textit{Rhole} and \textit{ftaper} parameters.  

\begin{figure*}
\includegraphics[width=0.96\textwidth]{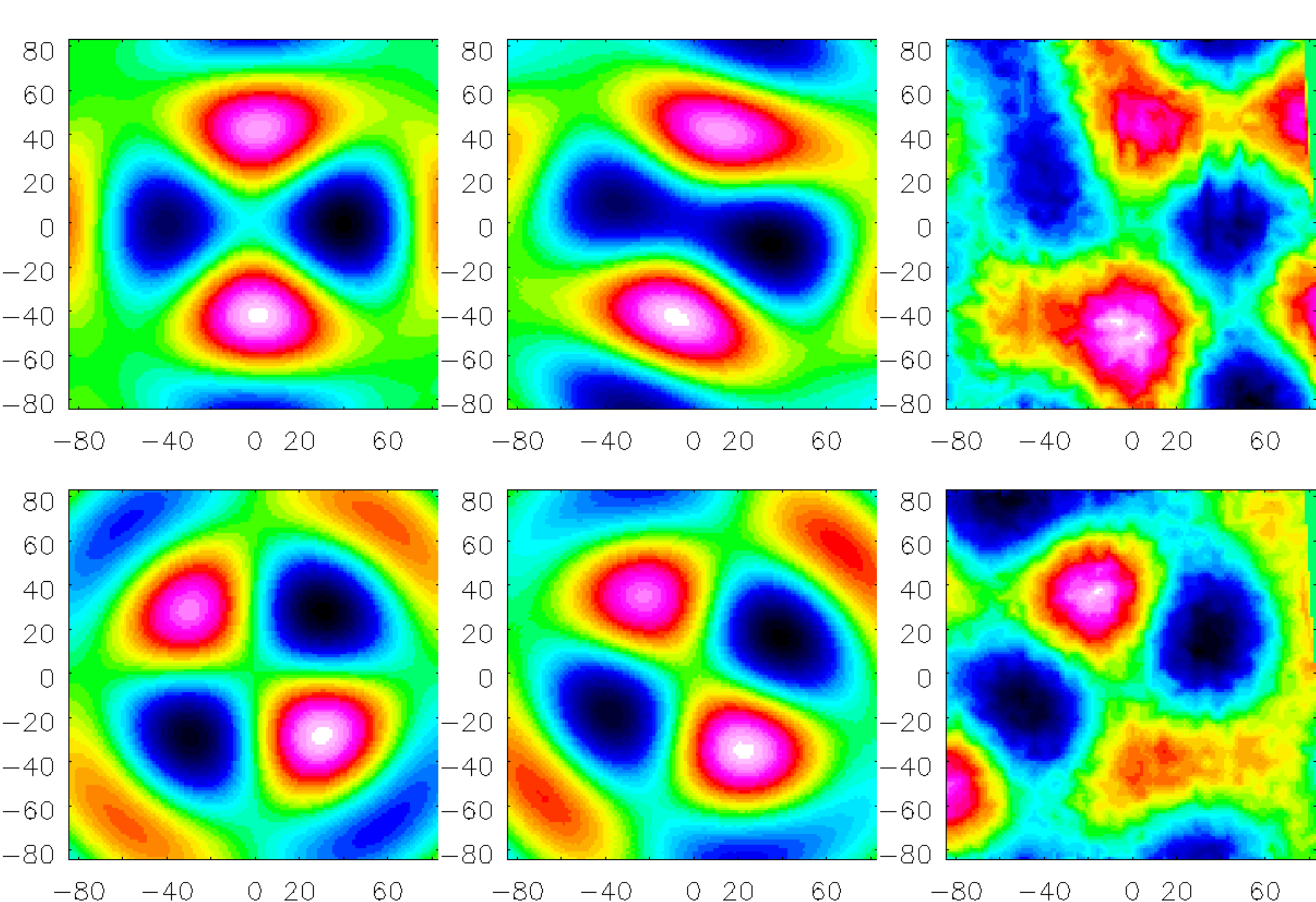}
\caption{ The panels of the figure shows the off-diagonal antenna Jones matrix element {\tt $R\leftarrow L$} of $E_i^{M} = \mathcal{F}^{-1}[\A_i^M]$, with the upper panels, show the real portion of $E_i^{M}$ and the lower panels show the imaginary part of $E_i^{M}$ at 1.013~GHz of antenna 27. The left most panels (upper and lower) $E_i^{M}$ that includes an optimized second order polynomial in phase. The figures in the center of the complex $E_i^{M}$, includes an $\approx18^{\circ}$ rotation in addition to the second order polynomial in phase. In the two panels on the right, the real and imaginary parts of the measured holographic $E_i^{M}$ is shown.}
\label{fig:rlcompare}
\end{figure*}

In Fig.~\ref{fig:rlcompare} the panels on the left represent the real
(upper left) and imaginary (lower left) parts of the off-diagonal
antenna Jones matrix element {\tt $E^{R\leftarrow L}$} of the
  model. The inclusion of a second order phase term in the antenna
alters the side-lobe flux but does not alter the general morphology of the
clover-leaf pattern as is the case in panels on the right in
Fig.~\ref{fig:rlcompare}. The real (upper right) and imaginary (lower
right) parts of the off-diagonal antenna Jones matrix element {\tt
  $E^{R\leftarrow L}$} of the measured holographic map. The
altered morphology of the lobes mimics a rotation of the VP. We
 therefore introduced rotation of the antenna VP as an additional
free parameter in the minimization which lead to a more realistic model VP shown in the center
panels. A rotation of $\approx18^{\circ}$ gave the least residuals
with respect to the holographic data.  A similar rotation is
  quite clearly seen in the polarization squint vector as well for all
  antennas and at bands in the holographic measurements.  The physical
  origin of this rotation is not yet understood.

\section{Conclusions}

The imaging performance of the \AWP\ is determined by our knowledge of the AIP. 
High dynamic range and high fidelity polarimetric imaging across wide-fields requires an extremely accurate understanding of the antenna AIP across the full bandwidth. The A-Solver approach of solving for the frequency-dependent AIP of antennas based on a parametrized model whose values are determined by comparison to holographic data is a viable approach to obtaining an accurate VP as demonstrated in this paper. The parameterized model captures the rapid frequency dependence of the AIP including the effects of standing waves. Modeling the central blockage as an apparent blockage in the model allowed for the accurate reconstruction of the amplitude of the VP side-lobe as a function of frequency. The parameterized model of the AIP is a naturally compact representation requiring fewer parameters to capture higher order frequency-dependent effects, than frequency-dependent modeling of antenna VP. 

An important point to note is that the product of the two AIPs
making the PB for each baseline is, in general, a complex valued
function, and not a purely real function as is assumed when imaging
without using the \AWP\ algorithm (the effective PB with \AWP\ is
$\sqrt{PB^M\cdot PB^{\circ^\dag}}$, which is real-valued at the level
the model $PB^M$ accurately models the real $PB^\circ$) .  This could
be due to differences between the two AIPs involved and/or
non-Hermitian structure of the AIPs due to various EM or antenna
structural effects.  The PB pattern is already quite complex, and as
discussed Sec.~\ref{Sec:MODELING}, directly modeling even the
real-valued PB is difficult, approximate and needs many more free
parameters. In addition to this, modeling of the complex valued PB
also has all the additional numerical complications involved in
directly fitting to any complex valued data.  In contrast, the
physical modeling approach described in this paper models the PB in
the aperture plane.  This not only requires significantly smaller
number of parameters, the parameters themselves are real-valued
describing the physics of the optics (here, via the antenna structural
parameters).  The fitting procedure, therefore, deals with real-valued
parameters.  This has significant numerical advantages, and computational advantages
in the production of parameterized convolution functions for a given 
frequency.

\begin{acknowledgements}
  This work was done using the R\&D branch of the CASA code base.  We
  thank R.~Perley for carrying out the holography and O.~Smirnov for carrying out
 various illuminating numerical experiments with the data. We thank
  James Robnett and Erik Bryer for their extensive assistance in deployment of 
  the minimization runs on AWS. Support for this work was provided by the NSF through the Grote Reber Fellowship Program administered by Associated Universities, Inc./National Radio Astronomy Observatory.The National Radio Astronomy Observatory is a facility of the National Science Foundation operated under cooperative agreement by Associated Universities, Inc.
\end{acknowledgements}

\software{CASA (McMullin et al. 2007)}

\end{document}